\documentclass[conference]{IEEEtran}
\usepackage{cite}
\usepackage{amsmath,amssymb,amsfonts}
\usepackage{lineno}
\usepackage{caption}
\usepackage{algpseudocode}
\usepackage{graphicx}
\usepackage{subcaption}
\usepackage{textcomp}
\usepackage{xcolor}

\usepackage[ruled,vlined,linesnumbered]{algorithm2e}
\usepackage{amsmath}
\usepackage[skip=0pt]{caption}

\def\BibTeX{{\rm B\kern-.05em{\sc i\kern-.025em b}\kern-.08em
    T\kern-.1667em\lower.7ex\hbox{E}\kern-.125emX}}
\begin{document}

\title{VersaSlot: Efficient Fine-grained FPGA Sharing with Big.Little Slots and Live Migration in \\ FPGA Cluster
}

\author{\IEEEauthorblockN{Jianfeng Gu, Hao Wang, Xiaorang Guo, Martin Schulz and Michael Gerndt}
\IEEEauthorblockA{Chair of Computer Architecture and Parallel Systems, Technical University of Munich \\ Munich, Germany \\ 
\{jianfeng.gu, whhao.wang, xiaorang.guo\}@tum.de, \{schulzm, gerndt\}@in.tum.de}
}

\maketitle

\begin{abstract}
As FPGAs gain popularity for on-demand application acceleration in data center computing, dynamic partial reconfiguration (DPR) has become an effective fine-grained sharing technique for FPGA multiplexing. However, current FPGA sharing encounters partial reconfiguration contention and task execution blocking problems introduced by the DPR, which significantly degrade application performance. In this paper, we propose VersaSlot, an efficient spatio-temporal FPGA sharing system with novel Big.Little slot architecture that can effectively resolve the contention and task blocking while improving resource utilization. For the heterogeneous Big.Little architecture, we introduce an efficient slot allocation and scheduling algorithm, along with a seamless cross-board switching and live migration mechanism, to maximize FPGA multiplexing across the cluster. We evaluate the VersaSlot system on an FPGA cluster composed of the latest Xilinx UltraScale+ FPGAs (ZCU216) and compare its performance against four existing scheduling algorithms. The results demonstrate that VersaSlot achieves up to 13.66x lower average response time than the traditional temporal FPGA multiplexing, and up to 2.19x average response time improvement over the state-of-the-art spatio-temporal sharing systems. Furthermore, VersaSlot enhances the LUT and FF resource utilization by 35\% and 29\% on average, respectively.

% The results demonstrate that VersaSlot achieves an average response time up to 13.66x lower than traditional temporal FPGA multiplexing, and up to 2.19x lower than the state-of-the-art spatio-temporal sharing system.
% we introduce an efficient slot allocation and scheduling algorithm and design a seamless cross-board switching and live migration mechanism to maximize FPGA multiplexing in the cluster

\end{abstract}

% \begin{IEEEkeywords}
% FPGA sharing, Partial Reconfiguration, Scheduling.
% \end{IEEEkeywords}

\section{Introduction}
Field Programmable Gate Arrays (FPGAs) have rapidly gained popularity in edge and data center computing, offering efficient application acceleration with higher flexibility and lower power consumption \cite{chung2018serving, wang2022via, wu2023fasda, zeng2023dfgas, zhao2023sagraph}. Leading cloud providers \cite{alibabacloudWhatFPGA,amazonAmazonInstances,microsoftMicrosoftAzure} have integrated FPGAs into their platforms, allowing users to develop highly customized and on-demand applications with Quality of Service. Meanwhile, modern FPGAs have significantly enhanced their capabilities with advanced manufacturing processes and increased logic elements, able to handle more complex and diverse computational demands \cite{bobda2022the,damiani2022blastfunction,ringlein2021acase}. These improvements facilitate fine-grained FPGA sharing \cite{ahmed2018amorphos,zha2020virtualizing,dhar2022dml,mandava2023nimblock} in multi-user and multi-task environments.

Traditional FPGA virtualization and sharing techniques \cite{amazonAmazonInstances,putnam2014reconfigurable,tarafdar2017enabling} allocate the entire FPGA to a single application and implement the time-multiplexing by performing a full fabric reconfiguration. This approach introduces significant context switch overhead and often leads to FPGA under-utilization, as applications may not require the full FPGA resources. Instead, \textit{Dynamic Partial Reconfiguration} (DPR) \cite{dpr2020document} enables an FPGA to split the programmable fabric into multiple \textit{slots}, which can be reconfigured independently to host arbitrary application logic at runtime. It provides the capability to perform \textit{Partial Reconfiguration} (PR) on a slot while other slots continue running. Currently, DPR has emerged as the most efficient and flexible technique for spatio-temporal FPGA sharing and virtualization \cite{zha2020virtualizing,dhar2022dml,mandava2023nimblock} in the data center. 

However, current DPR-based spatio-temporal FPGA sharing still contains numerous overlooked challenges, leading to significant application performance degradation. First, in the current FPGA Processing System (PS), the PR module, \textit{PCAP} (Processor Configuration Access Port), is limited to serial bitstream loading. This serial access causes frequent mutual blocking when multiple applications require PR concurrently. As the number of applications sharing an FPGA increases, the PR contention intensifies, resulting in highly prolonged and unpredictable application response times. Meanwhile, the mutual PR blocking further severely disrupts applications' original pipelines, leading application tasks with dependencies in various slots to wait for each other to complete, preventing further execution. Recent systems like Nimblock \cite{mandava2023nimblock} and DML\cite{dhar2022dml} overlook these problems, resulting in application performance that falls short of the expected pipeline outcomes. Moreover, these works \cite{mandava2023nimblock,dhar2022dml} used single-core task scheduling, in which PR operations further block the launching of application tasks, leading to severe task blocking problems.

Second, the uniform size of FPGA slots in previous work \cite{dhar2022dml,mandava2023nimblock} imposes strict constraints on task partitioning. For example, when optimizing application tasks using Xilinx's High-Level Synthesis (HLS) tools, resource consumption typically exhibits stepwise increases rather than linear growth, which can easily cause resource over-subscription and under-utilization within slots. While prior work \cite{ahmed2018amorphos} attempted to address the problem by dynamically adjusting slot sizes based on application resource demand, the solution needs regenerating bitstreams at runtime, incurring large runtime overhead.

 % Most applications usually show imbalanced use of different FPGA resources (eg. FFs, LUTs, and BRAMs), preventing full utilization of a slot.

To address these challenges, we propose \textbf{VersaSlot} (Versatile Slots), an efficient spatio-temporal FPGA sharing system, to achieve more effective FPGA multiplexing for applications. To mitigate PR contention among application tasks, VersaSlot introduces a novel \textit{Big.Little} slot architecture. The architecture allows the system to bundle and load multiple tasks simultaneously to a Big slot for internal execution at runtime, thereby eliminating further PR contention with tasks from Little slots. Within the Big slot, tasks can still maintain their parallel or serial pipelines, as they do in the Little slots, but without internal PR interference. Meanwhile, reduced PR contention also allows Little slots to load task bitstreams promptly, mitigating blocking delays and pipeline disruptions.

However, similar to ARM's big.LITTLE cores \cite{armDocumentationx2013}, the heterogeneous \textit{Big.Little} slot design introduces increased complexity for resource allocation and scheduling. As for resource allocation, we propose an adaptive algorithm that combines primary allocation and redistribution along with binding and rebinding. The algorithm enables applications to flexibly switch between using Big and Little slots at runtime while also maximizing the utilization of available slot resources. The complementary between Big and Little slots can help mitigate under-utilization and over-subscription problems. In the context of scheduling, the VersaSlot dynamically bundles 3-in-1 tasks at runtime and schedules the corresponding tasks to available slots. The 3-in-1 task bundling can enhance the flexibility in task partitioning of applications. Importantly, VersaSlot decouples the PR from scheduling logic and distributes them across two CPU cores for asynchronous execution, thus effectively eliminating the task execution blocking problem.

For specific applications with few tasks and large batch sizes, PR occurs infrequently. By dividing Big slots into multiple Little slots, more applications can efficiently share resources, making an FPGA with only Little slots (\textit{Only.Little}) also necessary. However, switching an FPGA between \textit{Only.Little} and \textit{Big.Little} configurations requires restarting the system and interrupting all task executions, which introduces a large runtime overhead. To address this, we propose a seamless cross-board switching and live migration mechanism based on system PR contention levels within a cluster, enabling the FPGA slot configuration switching with low overhead.

In a nutshell, the contributions are summarized as follows:
% \vspace{-0.5mm}
\begin{itemize}
\item We propose VersaSlot, an efficient spatio-temporal FPGA sharing system that introduces the novel \textit{Big.Little} slot architecture. VersaSlot addresses the significant PR contention problem inherent in FPGA fine-grained sharing. 

\item We design an efficient slot allocation and dual-core scheduling algorithm for the \textit{Big.Little} slot architecture to dynamically and efficiently utilize both kinds of slots and prevent the task execution blocking problem.

\item We propose a seamless cross-board switching and live migration mechanism, enabling applications to transition among FPGAs with different slot configurations in a cluster for more efficient execution with low overhead.

\item We evaluate the VersaSlot system on a real FPGA cluster, against four existing scheduling algorithms. VersaSlot achieves an average response time up to 13.66x lower than the traditional FPGA multiplexing, and up to 2.19x lower than the state-of-the-art system, and enhances the LUT and FF utilization by 35\% and 29\% on average.
\end{itemize}

\section{Background}
\label{sec:background}
% \vspace{-0.5mm}
\label{subsec:dpr_subsection}
Dynamic Partial Reconfiguration (DPR), also known as Dynamic Function eXchange (DFX) from AMD/Xilinx \cite{amdTechnicalInformation}, is an advanced FPGA configuration technique that enables dynamic reconfiguration of specific regions of the FPGA without interrupting the overall system operation. Instead, in traditional FPGA configuration flows, any functional modification requires reloading the entire bitstream file, leading to system downtime and a full restart, which causes large runtime overhead.  The DFX process relies on the Processor Configuration Access Port (PCAP) \cite{amdPcap} to load the partial bitstream from memory to the corresponding section of the FPGA. As it is essential to confirm the successful loading of the partial bitstream, DFX inherently prevents the PCAP from loading a new partial bitstream until the current one is fully loaded and blocks the corresponding CPU \cite{amdPcap}. Meanwhile, to avoid unpredictable impacts on the remaining FPGA circuitry during the loading phase, a dedicated IP, DFX Decouple, is employed to decouple the reconfigurable region from the rest of the circuit. Recently, the DFX has been extensively used in FPGA virtualization and sharing in the cloud \cite{zha2020virtualizing,dario2020doos,dhar2022dml,mandava2023nimblock}, but limited to serial partial bitstream loading.

\section{System Design}
% \vspace{-0.5mm}
\subsection{VersaSlot System Framework}
\vspace{-3mm}
\label{sec:versaslot_system_framework}
\begin{figure}[htbp]
  \centering
  \includegraphics[width=0.49\textwidth]{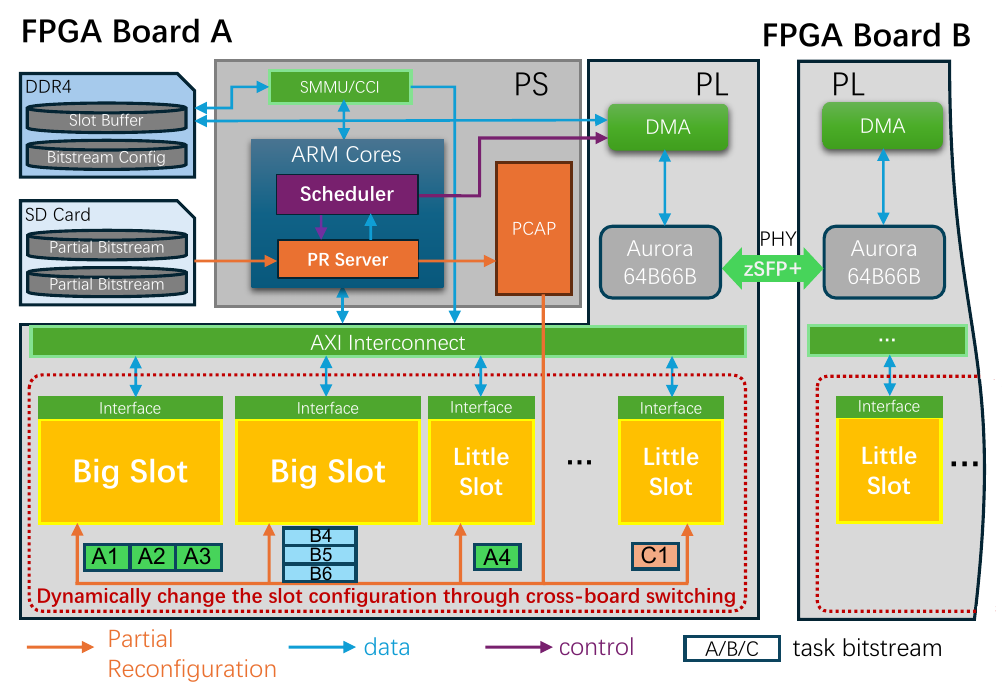}
  \caption{VersaSlot system with PS and PL in FPGA cluster.}
  \label{fig:versaslot_framework}

\end{figure}
% \vspace{-2mm}
The VersaSlot system, aligned with the typical structure of FPGA sharing design \cite{mandava2023nimblock,dhar2022dml}, consists of two main components: the embedded hypervisor within the Processing System (PS) and the FPGA fabric layout in the Programmable Logic (PL), as illustrated in Figure \ref{fig:versaslot_framework}. The VersaSlot hypervisor in the PS is implemented on ARM cores via bare-metal programming to minimize the system control overhead. Its primary responsibilities include scheduling and resource allocation, managing the loading of partial bitstreams and application data, and coordinating cross-board switching. For FPGA boards without a dedicated CPU, the hypervisor can run on the host CPU and control the FPGA via the PCIe interface. 

The VersaSlot hypervisor in the PS comprises two key modules: the scheduler and the PR server, both of which run as bare-metal applications on the ARM cores. At runtime, the scheduler allocates computational slot resources to each incoming application, updates the status of reconfigurable slots, and determines the next application to execute. The PR server, which handles partial reconfiguration, waits for PR requests from the scheduler. It loads pre-generated bitstreams from the SD card into memory and sends commands to the FPGA’s PCAP module to dynamically switch partial bitstreams for applications. However, the PCAP can only perform PR for one bitstream at a time and suspends the associated CPU during the process. Additionally, the PR server communicates status updates to the scheduler through the On-Chip Memory (OCM). The ARM cores utilize the System Memory Management Unit (SMMU) for address mapping and the AXI Interconnect to transfer application data to the FPGA slots for execution.

In the PL, based on the DPR technique, FPGA resources are divided into a \textit{static region} and partial reconfigurable Big and Little slots. The static region configures interfaces for the reconfigurable slots, enabling slots to communicate with the PS and memory via the AXI bus. The region can only be programmed once at system startup, while the reconfigurable parts can be dynamically reconfigured and switched to map different application logic at runtime. Before execution, application bitstreams are prepared offline in advance. Specifically, applications are partitioned into smaller tasks suitable for Little slots by synthesis resources via automated scripts. The \textbf{\textit{task}} represents a portion of the application and the basic execution unit for a slot. The VersaSlot system supports the dynamic \textbf{\textit{batch}} processing for applications. The fine granularity and batched structure of tasks facilitate an application to be organized in a pipeline with dependencies across slots during execution, which can increase parallelism, thus reducing execution time and enhancing resource utilization. A \textbf{\textit{slot}} refers to a reconfigurable region. Traditionally, the slots are designed to be uniform. In the VersaSlot system, we introduce a novel \textit{Big.Little} slot architecture, where three consecutive small tasks can be bundled as a big task for execution in a Big slot. In our system, an FPGA consists of either 2 Big slots and 4 Little slots (\textit{Big.Little}) or 8 Little slots (\textit{Only.Little}), but can be extended to any Big/Little configuration. The resource capacity of each Big slot is twice that of a Little slot. To ensure slot compatibility, the automated script generates partial bitstreams for each task adaptive to each slot. These generated bitstreams are stored on the SD card. Once partial reconfiguration is complete, the scheduler allocates buffers and launches the task.

The cross-board switching module is another important component in the PL, designed for board switching between different FPGA slot configurations. It utilizes the GT transceivers (zSFP+) to connect other FPGA boards. We introduce the Aurora IP core, an efficient communication protocol to transfer tasks, application information, and data directly via DMA to another FPGA unit when switching is activated.

\subsection{Big.Little Slots}
\label{subsec:big_littl_slots}
% \vspace{-1.0mm}
\begin{figure}[htbp]
   \captionsetup{skip=3pt}
  \centering
  \includegraphics[width=0.49\textwidth]{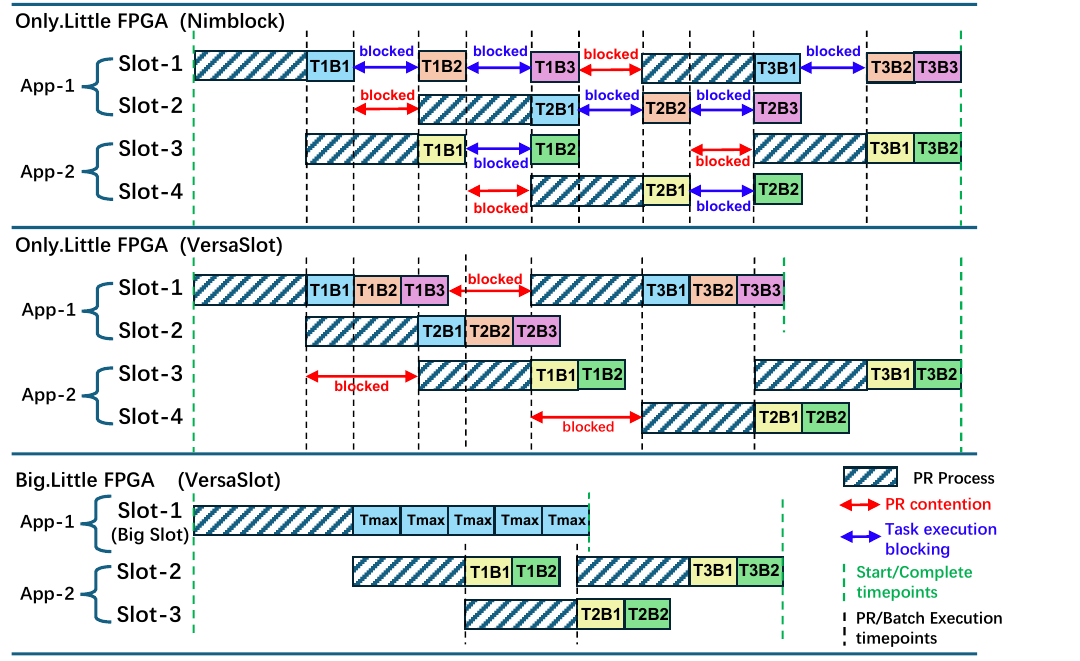}
  \caption{Versaslot with \textit{Big.Little} and \textit{Only.Little} dual-core scheduling alleviates the PR contention and task execution blocking problems, thus reducing application response time.}
  \label{fig:pr-blocked-example}
  \vspace{-1mm}
\end{figure}

\begin{figure}[htbp]
  \centering
  \includegraphics[width=0.49\textwidth]{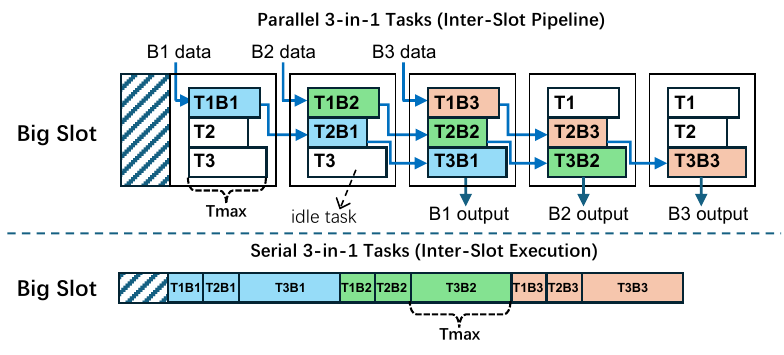}
  \vspace{-3mm}
  \caption{Parallel and serial bundling for the 3-in-1 task.}
  \label{fig:3in1-pipeline}
\end{figure}
\setlength{\textfloatsep}{1.5pt}

The \textit{Big.Little} architecture is a heterogeneous slot design for spatio-temporal FPGA sharing, coupling resource-intensive Big slots and standard-resource Little slots. Although Little slots can increase resource granularity for more applications to share an FPGA, they also bring severe mutual PR contention and task execution blocking problems due to inherent DPR limitation in serial bitstream loading and CPU blocking. For example, as shown in Figure \ref{fig:pr-blocked-example}, two applications (App-1 and App-2), each with 3 tasks (T1, T2, T3) and batch sizes of 3 (B1, B2, B3) and 2 (B1, B2), are processed on an FPGA with 4 Little slots, in which each application is allocated with 2 slots that forms the execution pipeline. In the Nimblock \cite{mandava2023nimblock}, after the App-1 completes the PR of its first task, the App-2 starts the PR of its first task, during which it blocks the bitstream loading of App-1's second task (T2) although App-1's first batch item of the first task (T1B1) has finished the execution. Subsequently, the PR of App-1's second task (T2) also blocks the PR of App-2's second task. Moreover, since Nimblock handles all operations and scheduling by a single CPU core, the PR process of App-2's T1 also blocks the task execution of App-1's T1B2, and the PR of App-1's T2 blocks the task execution of App-2's T1B2. As the two applications continue running, mutual PR contention and task execution blocking alternate, disrupting each application's original pipeline and significantly increasing response times. The figure shows only 4 slots with 2 applications, but as the number of shared slots and applications increases, this effect becomes more severe.

Within our uniform Little slots (\textit{Only.Little}) FPGA design, VersaSlot decouples PR logic from the scheduler to a dedicated PR server on a separate CPU core, which effectively resolves task execution blocking. As in Figure \ref{fig:pr-blocked-example}, while other applications perform PR operations, the scheduler can still promptly trigger batch execution of tasks without waiting for specific PR completion points. Additionally, we pre-load tasks of the same application into the slots to further enhance pipeline efficiency. The \textit{Only.Little} design is suited for specific applications with few tasks and large batch sizes, as PR occurs infrequently. But PR contention still persists in most workloads.

Therefore, we propose the Big slots, which can dynamically bundle and load 3 tasks (3-in-1 task) of an application simultaneously for internal execution at runtime. As shown in Figure \ref{fig:pr-blocked-example}, with VersaSlot's \textit{Big.Little} slots, once App-1 and its tasks are bundled and scheduled to execute on the Big slot, App-1 only needs to trigger tasks' batch execution without frequent PR loading, thus avoiding outside PR blocking from Little slots. Meanwhile, App-2's tasks can complete its pipeline smoothly without the PR blocking from App-1. Thus, the PR contention of applications in both Big and Little slots is alleviated. Furthermore, as in Figure \ref{fig:3in1-pipeline}, the tasks within Big slots can be organized in a pipeline as well. Since each batch’s parallel execution time on the Big slot equals the longest of the three tasks ({\small$T_{\text{max}}$}), serial execution is preferable when {\small$T_{\text{max}} \cdot (N_{\text{batch}} + 2) > \sum (T_1 + T_2 + T_3) \cdot N_{\text{batch}}$}. With this criterion, the VersaSlot system can select the optimal 3-in-1 task bitstream for execution at runtime. We set the bundling size to be 3 based on the Big slot's resource capacity to accommodate tasks and its fewer idle task cycles in pipelines than a larger size.

\subsection{Slot Allocation and On-board Scheduling}
% \vspace{-1.1mm}
Similar to the ARM big.LITTLE cores \cite{armDocumentationx2013}, the heterogeneous \textit{Big.Little} slot architecture introduces greater complexity in resource allocation and scheduling. We propose an efficient slot allocation and scheduling algorithm for the architecture. When an application {\small$A_i$} enters the candidate list, it is first added to the waiting list {\small$C_{wait}$} for slot allocation. Once the application acquires slot resources {\small$R_{A_i}$} (the maximum number of slots to use), all of its tasks are pushed into the ready list {\small$Q_T$}, awaiting scheduling for execution in Big or Little slots. 

\subsubsection{Slot Allocation}
\begin{algorithm}[hbt]
\caption{Slot Allocation}
\small
\label{alg:slot_allocation}
% Set input/output keywords
\SetKwInput{KwIn}{Input}
% \SetKwInput{KwOut}{Output}
% Set other keywords
\SetKwComment{Comment}{/* }{ */}
\SetKwIF{If}{ElseIf}{Else}{if}{then}{else if}{else}{end if}
% \SetKwFor{For}{for}{do}{end for}
% \SetKwFor{While}{while}{do}{end while}

\KwIn{
    $C_{wait} = \{A_i\}$: apps ${A_i}$ waiting for slot allocation; \\
    $S_{Big} = \{A_i\}$: the list of apps allocated with Big slots; \\ 
    $S_{Little} = \{A_i\}$: the list of apps allocated with Little slots; \\
     $B_{avail}$: idle Big slots;  \, $L_{avail}$: idle Little slots; (number);\\ 
    $Q_T$: the list of ready tasks waiting for scheduling; \\
    $N_{T_{A_i}}$: the number of unfinished ready tasks of $A_i$; \\
    $O_{A_i} = (O_{A_i}^{B}, O_{A_i}^{L})$: optimal Big/Little slots for $A_i$'s pipeline 
}
% % Algorithm outputs
\KwOut{
     $R_{A_i} = (R_{A_i}^{B}, R_{A_i}^{L})$: Big/Little slots allocated to $A_i$ \\
} 
% Algorithm main body
% 
$B_{avail} = B_{total} - \sum N_{T_{A_i}}, \, A_i \in S_{Big} \land T_{A_i} \in Q_T $; \\
% {\footnotesize$B_{avail} = B_{total} - \sum N_{T_{A_i}}, \, A_i \in S_{Big} \land T_{A_i} \in Q_T $;}  \\
{\footnotesize\tcp{No slot available, skip the allocation.}}
\If{$B_{avail} \leq 0$ and $L_{avail} \leq 0$}{
    return; 
}

{\footnotesize\tcp{Unbind apps with Little slots for rebinding.}}
\ForEach{$A_i \in S_{Little}$ \textbf{and} $B_{avail} > 0$}{
    \If{not isAppStarted($A_i$)}{
        $S_{Little} = S_{Little}\setminus \{A_i\}$; \, $C_{wait} = C_{wait} \cup \{A_i\}$;
    }
}

{\footnotesize\tcp{Primary allocation.}}
{\footnotesize $L_{left} = L_{total} - \sum \min(R_{A_i}^L, N_{T_{A_i}}), A_i \in S_{Little} \land T_{A_i} \in Q_T$;}
\ForEach{$A_i \in C_{wait}$}{
    {\footnotesize\tcp{Binding, prioritize big slot allocation.}}
    \If{$B_{avail} > 0$ \textbf{and} canBundle($A_i$)}{
        $R_{A_i} = (O_{A_i}^{B}, 0)$ ; $S_{Big} = S_{Big} \cup \{{A_i}\}$;\\
        $B_{avail} = B_{avail} - 1$; \textbf{continue};
    }
    {\footnotesize\tcp{Binding, allocation with Little slots.}}
    \If{$L_{avail} > 0$ and $L_{left} > 0$}{
        $R_{A_i} = (0, O_{A_i}^{L})$; $S_{Little} = S_{Little} \cup \{{A_i}\}$;\\
        $L_{left} = L_{left} - R_{A_i}^{L}$; \\
    }
}

{\footnotesize\tcp{Redistribution for left Little slots.}}
\If{$L_{left} > 0$}{
    \ForEach{$A_i \in S_{Little}$ \textbf{and} $L_{left} > 0$}{
        $\delta = N_{T_{A_i}} - O_{A_i}^{L}, \, T_{A_i} \in Q_T $;  \\
        $R_{A_i} = (0, \min(L_{left}+R_{A_i}^{L}, \delta + R_{A_i}^{L}))$; \\
        $L_{left} = L_{left} - \delta$;
    }
} 
\Return{$\{R_{A_i}\}$, \text{updated} $S_{Big}$, $S_{Little}$, $C_{wait}$}\;
\end{algorithm}

\setlength{\textfloatsep}{0mm}

As detailed in Algorithm \ref{alg:slot_allocation}, slot allocation is based on available slots and system status, featuring the processes of primary allocation and redistribution along with binding and rebinding. In the primary allocation, the system first allocates slots for applications in the waiting list $C_{wait}$. It prioritizes allocating Big slots to applications that can bundle tasks, as Big slots can significantly mitigate PR contention. After the Big slot allocation, Little slots are assigned to applications based on the optimal slot count {\small$O_{A_i}^{L}$}. The optimal slot allocation, derived through integer linear programming (ILP) as in \cite{dhar2022dml,mandava2023nimblock}, represents the most efficient slot configuration for pipeline execution within either Little or Big slots, whose value is usually lower than the task count. The primary allocation ensures applications execute promptly along their task pipelines. When additional Little slots are left, Versaslot performs the redistribution based on applications already assigned with Little slots, prioritizing those at the front of the runnable queue to obtain the maximum needed slots. This redistribution effectively avoids slot idling and maximizes overall slot utilization. Meanwhile, to prevent situations where Big slots are idle while Little slots are overburdened, we propose a rebinding method for load balancing. For applications assigned to Little slots but not yet executing, we unbind them for new slot allocation, returning them to a waiting state. This ensures that Big slots can receive new applications for execution. But applications bound to the big slots can only complete all their tasks in the Big slots to avoid Big slot blocking caused by task dependencies.

\subsubsection{On-board Scheduling}
\begin{algorithm}[hbt]
\caption{Scheduling Algorithm}
\small
\label{alg:scheduling}
% Set input/output keywords
\SetKwInput{KwIn}{Input}
% \SetKwInput{KwOut}{Output}
% Set other keywords
\SetKwComment{Comment}{/* }{ */}
\SetKwIF{If}{ElseIf}{Else}{if}{then}{else if}{else}{end if}
% \SetKwFor{For}{for}{do}{end for}
% \SetKwFor{While}{while}{do}{end while}

\KwIn{
    $S_{Big} = \{A_i\}$, $S_{Little} = \{A_i\}$, $\{R_{A_i}\}$; \\
    $B_{avail}$: idle Big slots;  \, $L_{avail}$: idle Little slots; (number); \\
    $Q_T$: the list of ready tasks waiting for scheduling; \\
    $U_{A_i}=(U_{A_i}^{B}, U_{A_i}^{L})$: number of Big/Little slots used by $A_i$;
}
% Algorithm main body
{\footnotesize\tcp{Add new tasks to the ready task list.}}
\ForEach{$A_i \in S_{Big} \cup S_{Little}$}{
    \If{$T_{A_i} \notin Q_T$}{
        $Q_T$.push\_back($T_{A_i}$); \\
    }
}
{\footnotesize \tcp{Bundle 3-in-1 task for Big slots online.}}
\ForEach{$A_i \in S_{Big}$ \textbf{and} $\{T_{A_i}\} \subset Q_T $}{
    \If{\textbf{not} isBundled($\{T_{A_i}\}$)}{
        $T_{A_i}^{bundled}$ = $3in1Bundle(\{T_{A_i}\})$; \\
        $Q_T$ replaces $\{T_{A_i}\}$ with $\{T_{A_i}^{bundled}\}$;
    }
}
{\footnotesize\tcp{Launch batch execution for running tasks.}}
\ForEach{$T_{A_i} \in Q_T$}{
    \If{isRunning$(T_{A_i})$ \textbf{and} waitBatchExec$(T_{A_i})$}{
        $launch(T_{A_i})$; \\
    }\ElseIf{isFinished$(T_{A_i})$}{
        $Q_T = Q_T \setminus T_{A_i}$; $isFinished(A_i)$$ \Rightarrow S = S \setminus A_i$;
    }
}
{\footnotesize\tcp{Schedule tasks to corresponding slots.}}
\ForEach{$T_{A_i} \in Q_T$}{
    \If{$A_i \in S_{Big}$ \textbf{and} $U_{A_i}^{B} < R_{A_i}^{B} $ \textbf{and} $B_{avail} > 0$}{
        PR$(T_{A_i})$ to an idle Big slot (Async in PR server); \\
        $T_{A_i}.running = True$; \, $B_{avail} = B_{avail} - 1$; 
        
    }
    \If{$A_i \in S_{Little}$ \textbf{and} $U_{A_i}^{L} < R_{A_i}^{L} $ \textbf{and} $L_{avail} > 0$}{
        PR$(T_{A_i})$ to an idle Little slot (Async in PR server); \\
        $T_{A_i}.running = True$; \, $L_{avail} = L_{avail} - 1$; 
        
    }
}
\end{algorithm}

\setlength{\textfloatsep}{0mm}

Algorithm \ref{alg:scheduling} outlines the overall scheduling process, which primarily includes 3-in-1 task online bundling, task execution launching, and scheduling tasks to their designated slots for PR. For newly added tasks in the ready list {\small$Q_T$} assigned with Big slots, the scheduler bundles them into serial or parallel 3-in-1 tasks at runtime and replaces them with the bundled tasks in the ready list for execution. Meanwhile, the scheduler updates task execution statuses, loads task data, and launches batch executions. In previous work \cite{dhar2022dml,mandava2023nimblock}, this step was often blocked by subsequent PR processes, leading to the task execution blocking problem. However, VersaSlot employs dual-core scheduling, which decouples the PR process and scheduling logic onto two separate CPU cores. When the scheduler assigns a new task to its corresponding available slot for PR, it sends an asynchronous request to the PR server without waiting for PR completion. Meanwhile, the scheduler limits application slot usage to its allocation, ensuring efficient spatio-temporal FPGA sharing. In VersaSlot's \textit{Only.Little} system, the scheduler follows the preemption mechanism as in \cite{mandava2023nimblock}, to prevent long-running tasks from monopolizing the FPGA. But the preemption brings more PR operations. The \textit{Big.Little} applies preemption only in Little slots since an application cannot occupy both Big and Little slots simultaneously. Meanwhile, the inherent design of the slot redistribution in \textit{Big.Little} also avoids monopolization.

\subsection{Cross-Board Switching and Live Migration}
% \vspace{-1.4mm}
\subsubsection{Live Migration}
As mentioned in section \ref{subsec:big_littl_slots}, \textit{Only.Little} and \textit{Big.Little} configurations have their own advantages for different application workloads, requiring an FPGA to switch between them to maximize performance. However, configurations related to slot size and interfaces are located in the \textit{static region}, which can only be programmed once during system start-up. Cross-board switching enables seamless transitions between configurations without rebooting, achieved through live workload migration. When switching is triggered, the original FPGA stops executing new tasks. Applications and tasks in the ready list, along with their buffers, are transferred via DMA to a new pre-configured FPGA. Once transferred, the new FPGA resumes task execution and processes upcoming new workloads. Meanwhile, ongoing tasks on the original FPGA continue to completion to avoid bitstream reloading overhead, and the FPGA is freed afterward to prevent excess resource usage. Therefore, in a cluster, a single available FPGA can enable cross-board switching for the entire system.

\subsubsection{Performance Degradation Metric {\small $D_{\text{switch}}$}}
To determine the optimal timing for the switching and evaluate system performance degradation from PR contention, we propose an evaluation metric, {\small$D_{\text{switch}}$}, considering all influencing factors.

% \vspace{-2mm}
{\small
\begin{equation}
    D_{\text{switch}} = \frac{N_{\text{blocked\_tasks}}}{N_{\text{PR}}} \times \frac{N_{\text{apps}}}{N_{\text{batch}}}, \quad (0 < D_\text{switch} < 1) 
\end{equation}}
where {\small $N_{\text{PR}} = \sum_{A_i} N_{T_{A_i}}, \ A_i \in R_c \cup R_s; \ N_{\text{batch}}=\sum_{A_i} B_{A_i}, \ A_i \in C$}.
The metric {\small$ D_{\text{switch}}$} combines the ratio {\small$\frac{N_{\text{blocked\_tasks}}}{N_{\text{PR}}}$} to evaluate current PR contention degree among tasks, and the ratio {\small$\frac{N_{\text{apps}}}{N_{\text{batch}}}$} to assess potential conflicts from candidate applications. It is normalized from 0 to 1, where higher values indicate more severe contention and less predictable application response time. The metric is recalculated after every $n$ updates to the application candidates queue with total completed ({\small$R_c$}) / running ({\small$R_s$}) / ready ({\small$R_r$}) applications. The {\small$N_{\text{blocked\_task}}$} represents the number of tasks blocked by PR contention during this period. The {\small$N_{\text{PR}}$} denotes the number of PR tasks {\small$\{T_{A_i}\}$} by all completed and running applications {\small$A_i$}. The {\small$N_{\text{blocked\_tasks}}$} is positively correlated with {\small$N_{\text{PR}}$}, which indicates a higher number of PR typically results in more blocked tasks and an increasing {\small$D_{\text{switch}}$}. Another two influential factors, batch size {\small$B_{A_i}$} and the number of applications {\small$N_\text{apps}$}, are significantly factored into the future contention evaluation. The {\small$N_{\text{batch}}$} is the total batch size of all applications {\small$A_i$} in the current application candidate queue {\small$C$}. If each application is allocated only one slot with batch size to be one, {\small$N_{\text{batch}} = N_{\text{apps}}$}, this indicates the worst-case scenario for PR conflicts and corresponds to the maximum value of {\small$D_{\text{switch}}$}.  

% \vspace{-0.5mm}
\begin{figure}[htbp]
  \centering
  \vspace{-2mm}
  \includegraphics[width=0.43\textwidth]{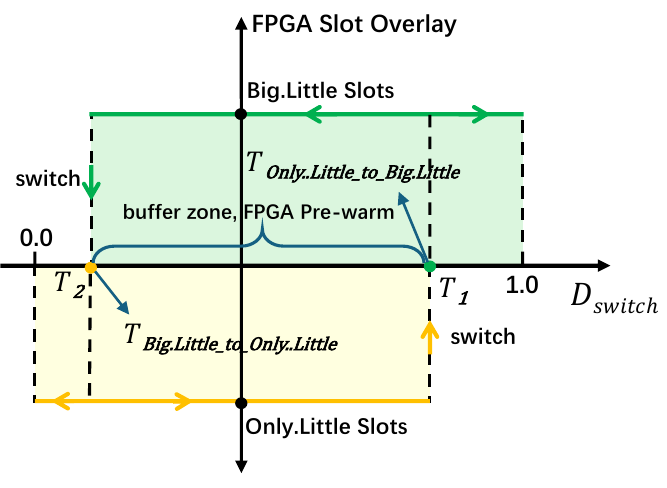}
  \caption{Cross-board switching loop with the buffer zone.}
  \label{fig:cross_board_switching_point}
\end{figure}
% \vspace{-3mm}

Inspired by Schmitt Trigger \cite{wikipediaSchmittTriggerWikipedia}, we propose a switch loop for seamless cross-board switching across multiple FPGAs. As shown in Figure \ref{fig:cross_board_switching_point}, as the {\small$D_{\text{switch}}$} increases and reaches the threshold  {\small$T_1$}, the system triggers the switching from the \textit{Only.Little} slots to the \textit{Big.Little} slots, thereby alleviating the PR contention. Similarly, when the {\small$D_{\text{switch}}$} decreases to the threshold {\small$T_2$}, it triggers the switching from the \textit{Big.Little} FPGA back to \textit{Only.Little} FPGA, allowing more future applications to share an FPGA. The two user-configurable thresholds adjust switching sensitivity, with a buffer zone to prevent frequent switching and ensure system stability. Meanwhile, when the {\small$D_{\text{switch}}$} metric enters the buffer zone, the system anticipates the direction of change, pre-configures the potential FPGA configuration, and loads task bitstreams into SD storage in a new FPGA. The pre-warming design further ensures seamless live migration with low overhead.

\section{Experiment and Evaluation}
% \vspace{-0.5mm}
The VersaSlot system was evaluated on an FPGA cluster with two latest Xilinx UltraScale+ FPGAs (ZCU216). We used the same benchmark as previous work \cite{mandava2023nimblock} and partitioned applications into multiple tasks based on the optimal fit between slot resources and task resource usage after synthesis. The applications include 3D Rendering (3DR, 3 tasks), LeNet (6 tasks), Image Compression (IC, 6 tasks), AlexNet (AN, 6 tasks), and Optical Flow (OF, 9 tasks). The partitioning and task bitstream generation were performed automatically by a TCL script in Vivado 2024.1. We compared the VersaSlot system with four existing scheduling algorithms/systems, including the traditional exclusive temporal multiplexing (Baseline) \cite{amazonAmazonInstances,putnam2014reconfigurable}, First-come-first-served spatio-temporal sharing (FCFS), round-robin (RR) from \cite{dario2020doos}, and the state-of-the-art spatio-temporal sharing system Nimblock \cite{mandava2023nimblock}. To simulate different congestion conditions, we randomly generated application workloads (10 sequences, 20 apps/sequence) with random batch size (5-30) and arrival intervals, including Loose (5000ms), Standard (1500ms-2000ms), Stress (150ms-200ms), and Real-time (50ms) for the evaluation. 
% \vspace{-1mm}
\subsection{Response Time Reduction and Tail latency}
% \vspace{-5mm}
\begin{figure}[htbp]
  \centering
  \includegraphics[width=0.49\textwidth]{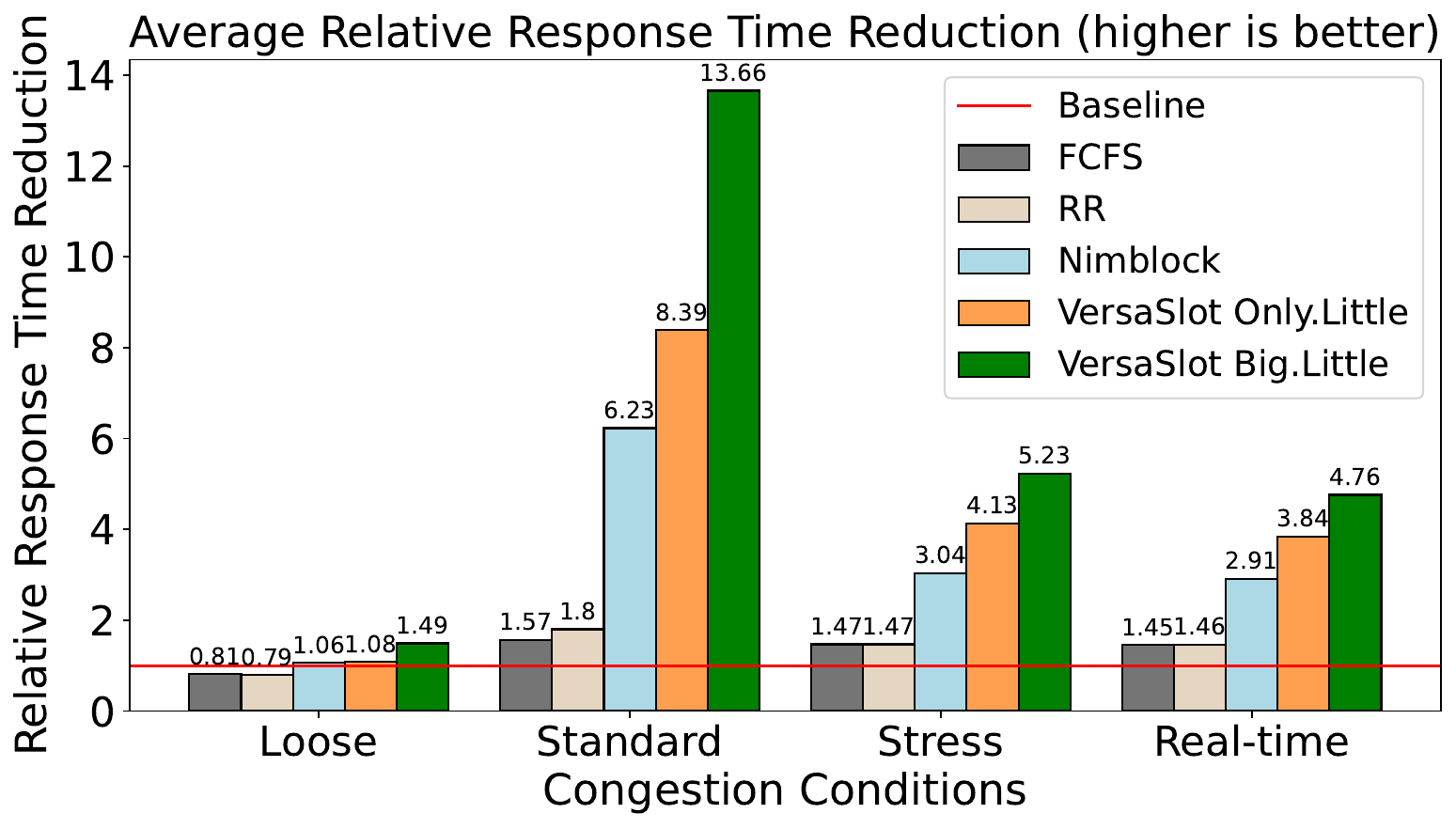}
  \caption{Relative response time reduction under different congestion conditions, normalized to the baseline.}
  \label{fig:result_response_time_reduction}
\end{figure}
\begin{figure}[htbp]
  \centering
  \includegraphics[width=0.49\textwidth]{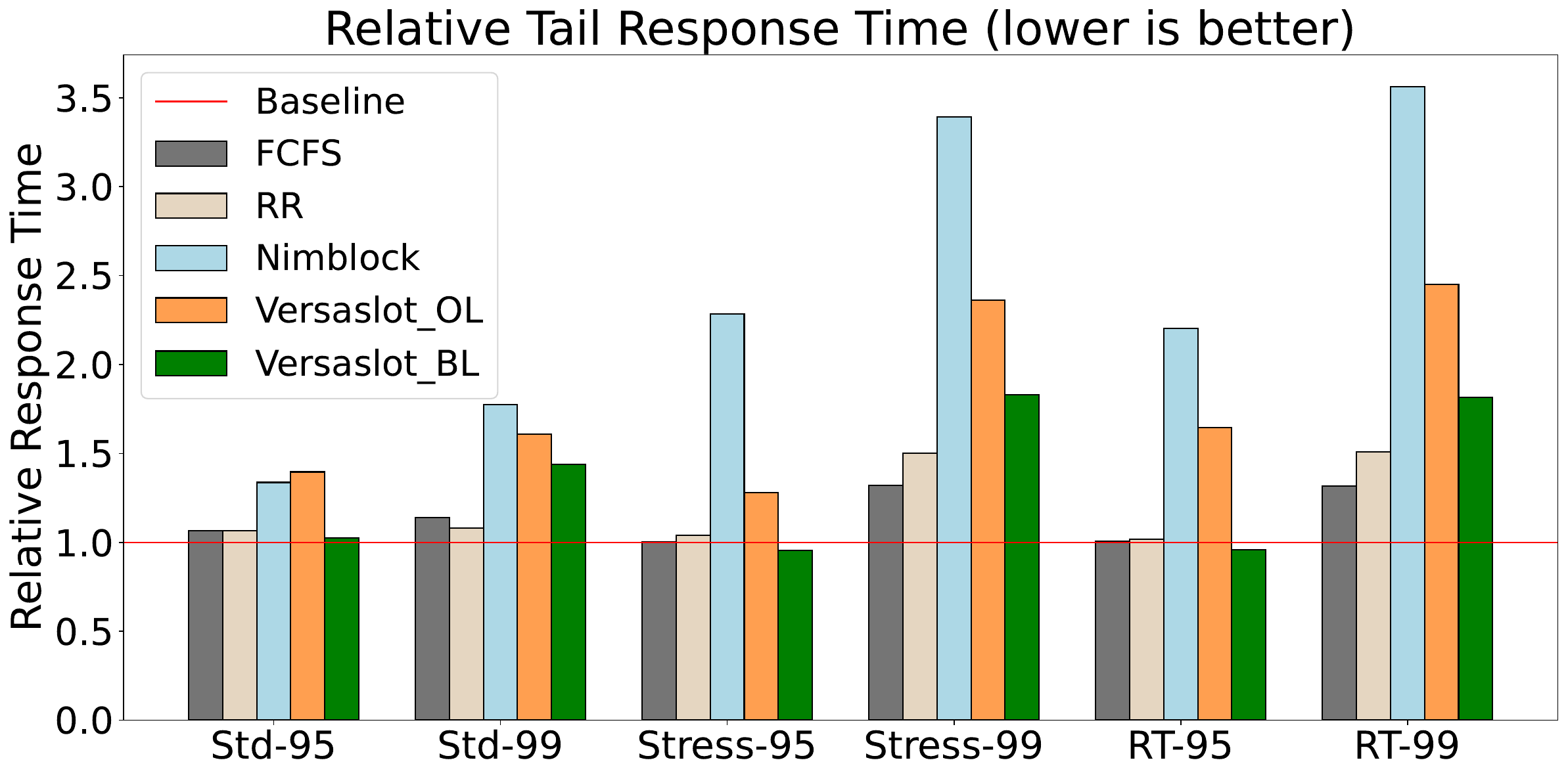}
  \caption{Tail response time normalized to the baseline.}
  \label{fig:tail_response_time}
\end{figure}

% \vspace{-3mm}
As shown in Figure \ref{fig:result_response_time_reduction}, we analyze the relative response time reduction under different congestion conditions. The VersaSlot \textit{Big.Little} design outperforms all other methods across all congestion conditions. Under standard workloads, it outperforms the baseline up to 13.66x. The FCFS and Round-robin perform slightly better than the baseline. Compared to the state-of-the-art system, Nimblock, the \textit{Big.Little} design reduces average response time up to 2.17x in standard workloads, 1.72x under stress, and 1.63x in real-time conditions, showcasing the efficiency of \textit{Big.Little} architecture and dual-core scheduling in fine-grained spatio-temporal FPGA sharing. Additionally, compared to the VersaSlot \textit{Only.Little} configuration, \textit{Big.Little} achieves average response time improvements by 63\%, 27\%, and 24\%, respectively, verifying its effectiveness in alleviating PR contention. Meanwhile, as shown in Figure \ref{fig:tail_response_time}, we further analyze the tail latency of applications, and the \textit{Big.Little} design consistently outperforms Nimblock in P95 and P99 metrics (95th and 99th percentiles) across all congestion conditions. Under stress workloads, \textit{Big.Little} improves these metrics by 83\% and 46\%, respectively, and by 56\% and 48\% under real-time workloads. While \textit{Big.Little} shows a slight increase in P99 tail latency compared to the baseline, it maintains or even improves P95 performance. Given its substantial advantage in response time, \textit{Big.Little} obviously demonstrates superior application performance than the baseline.

\begin{figure}[htbp]
% \vspace{-4mm}
  \centering
  \begin{subfigure}[b]{0.23\textwidth}
    \includegraphics[width=\textwidth]{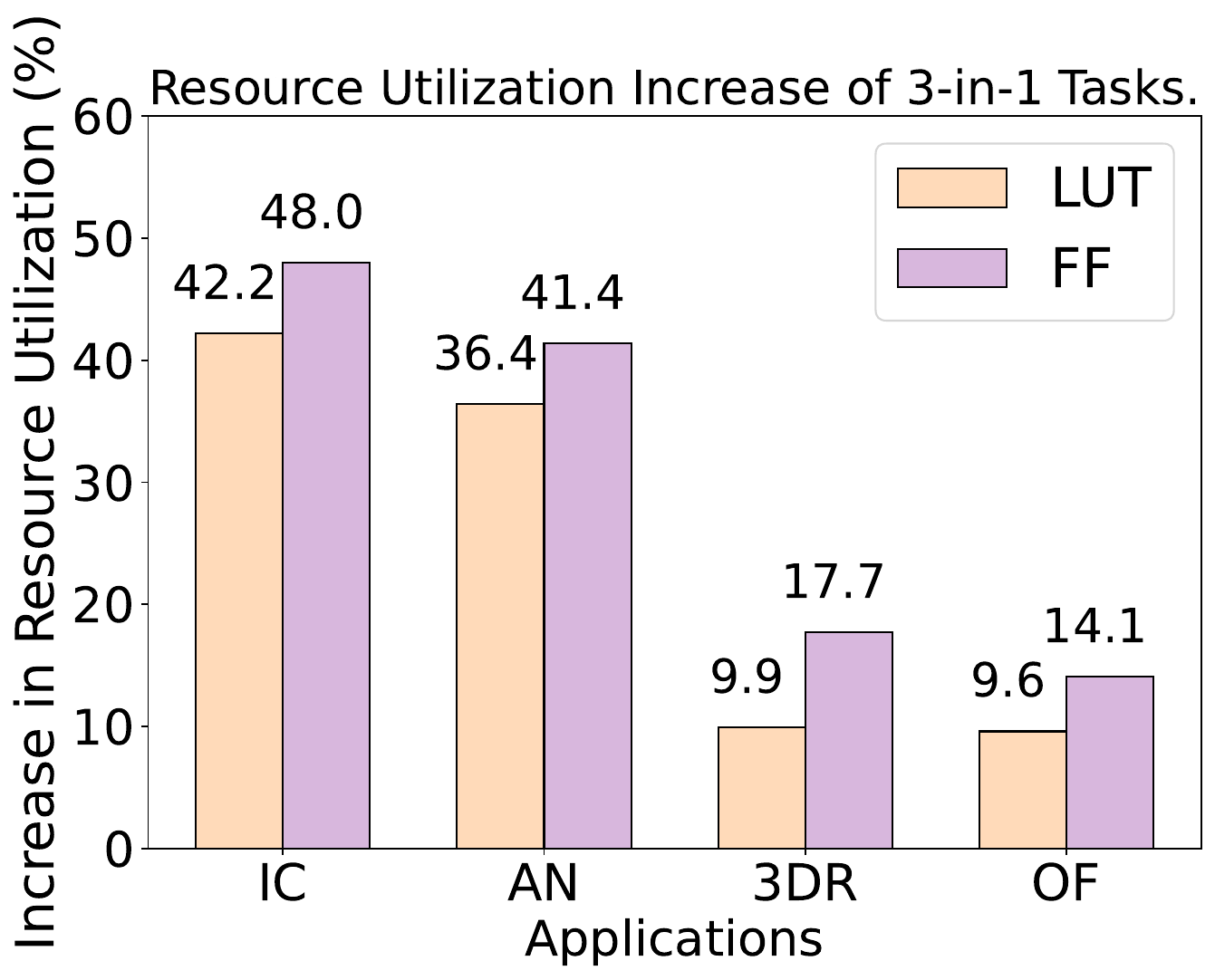}
  \end{subfigure}
%   \quad % or any other spacing command you like
  \begin{subfigure}[b]{0.24\textwidth}
    \includegraphics[width=\textwidth]{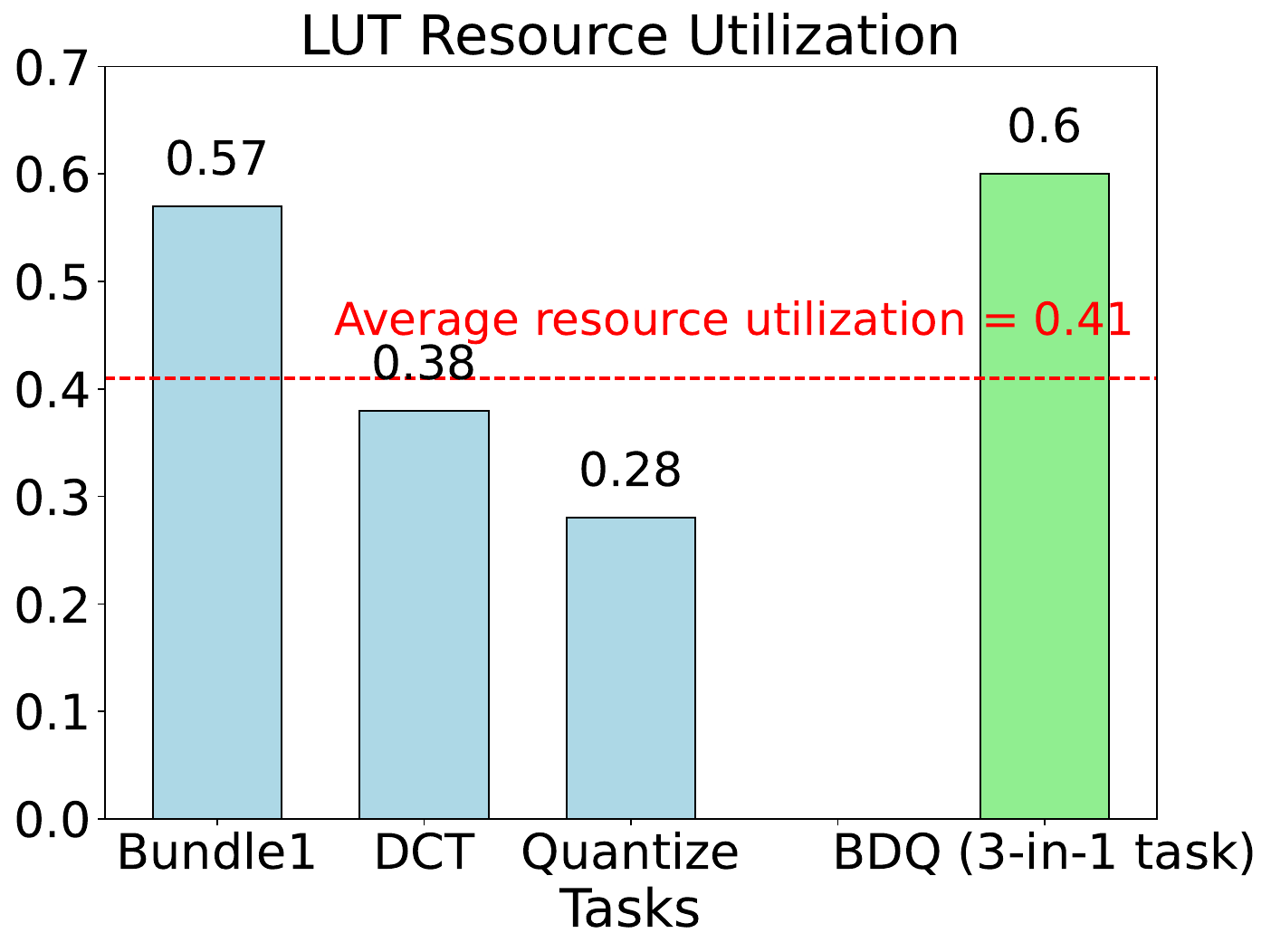}
  \end{subfigure}
  \caption{ Resource utilization improvement by 3-in-1 tasks.}
  \label{fig:resource_utilization}
\end{figure}
% \vspace{-1.5mm}
\subsection{Resource Utilization}
\vspace{-1mm}
Figure 7 shows the resource utilization improvement of 3-in-1 tasks in Big slots compared to running all tasks only in Little slots. All applications show significant LUT and FF utilization increases, with average improvements of 35\% and 29\%. Specifically, the right figure displays the LUT usage for the first three tasks of application IC and their bundled 3-in-1 task. Since task partitioning is normally based on synthesis resource usage, Bundle1’s LUT utilization decreases from 0.98 in synthesis to 0.57 in implementation. But, with bundling, the 3-in-1 task improves average utilization from 0.41 to 0.6.
% \vspace{-1mm}
\subsection{Cross-board Switching}
% \vspace{-5mm}
% \begin{figure}[htbp]
%   \centering
%   \includegraphics[width=0.42\textwidth]{results/cross_board_switching.pdf}
%   \caption{{\small$D_{\text{switch}}$} variation with cross-board switching.}
%   \label{fig:cross_board_switch_result}
% \end{figure}
% \vspace{-2.4mm}

\begin{figure}[htbp]
% \vspace{-4.5mm}
  \centering
  \begin{subfigure}[b]{0.33\textwidth}
    \includegraphics[width=\textwidth]{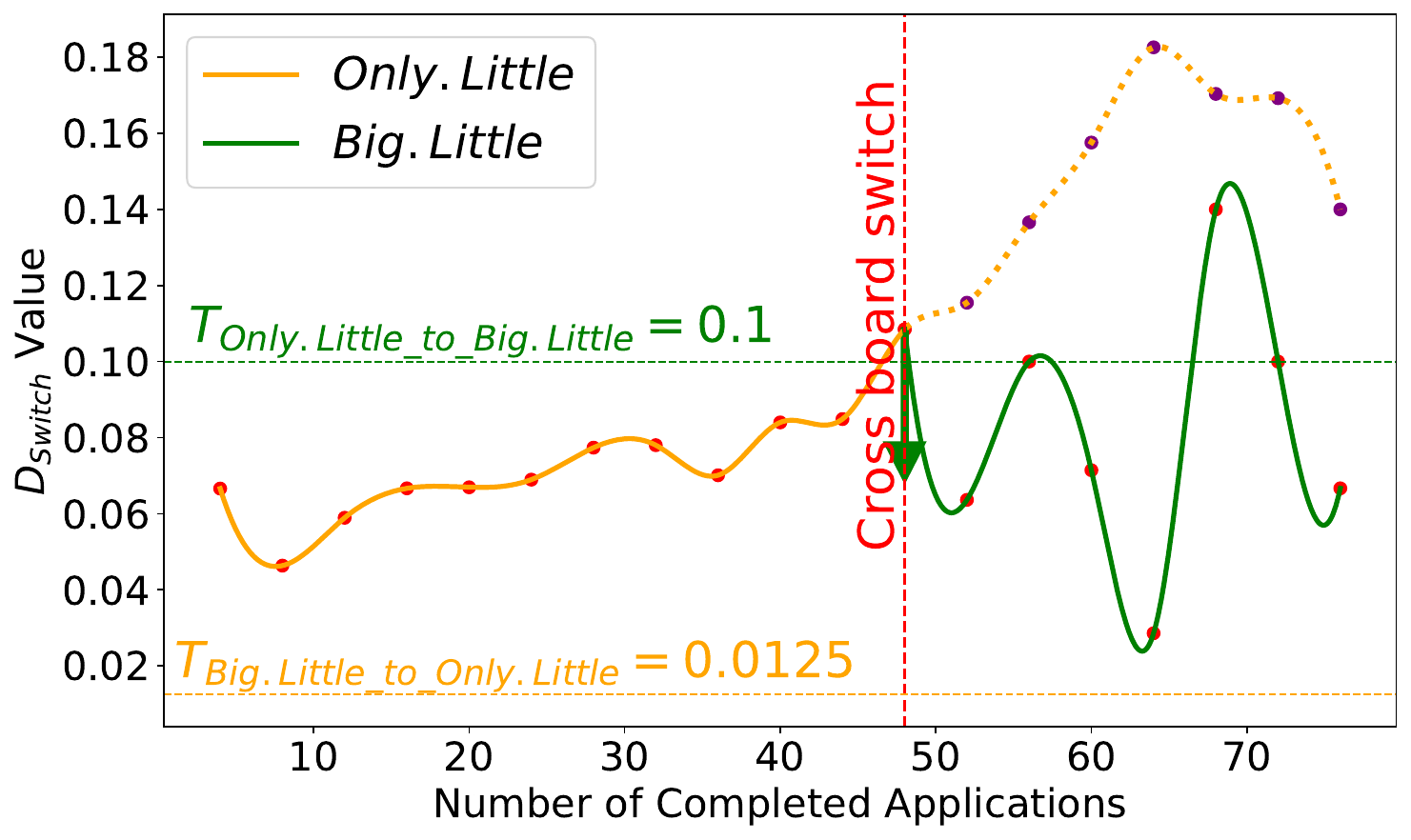}
  \end{subfigure}
  \begin{subfigure}[b]{0.15\textwidth}
    \includegraphics[width=\textwidth]{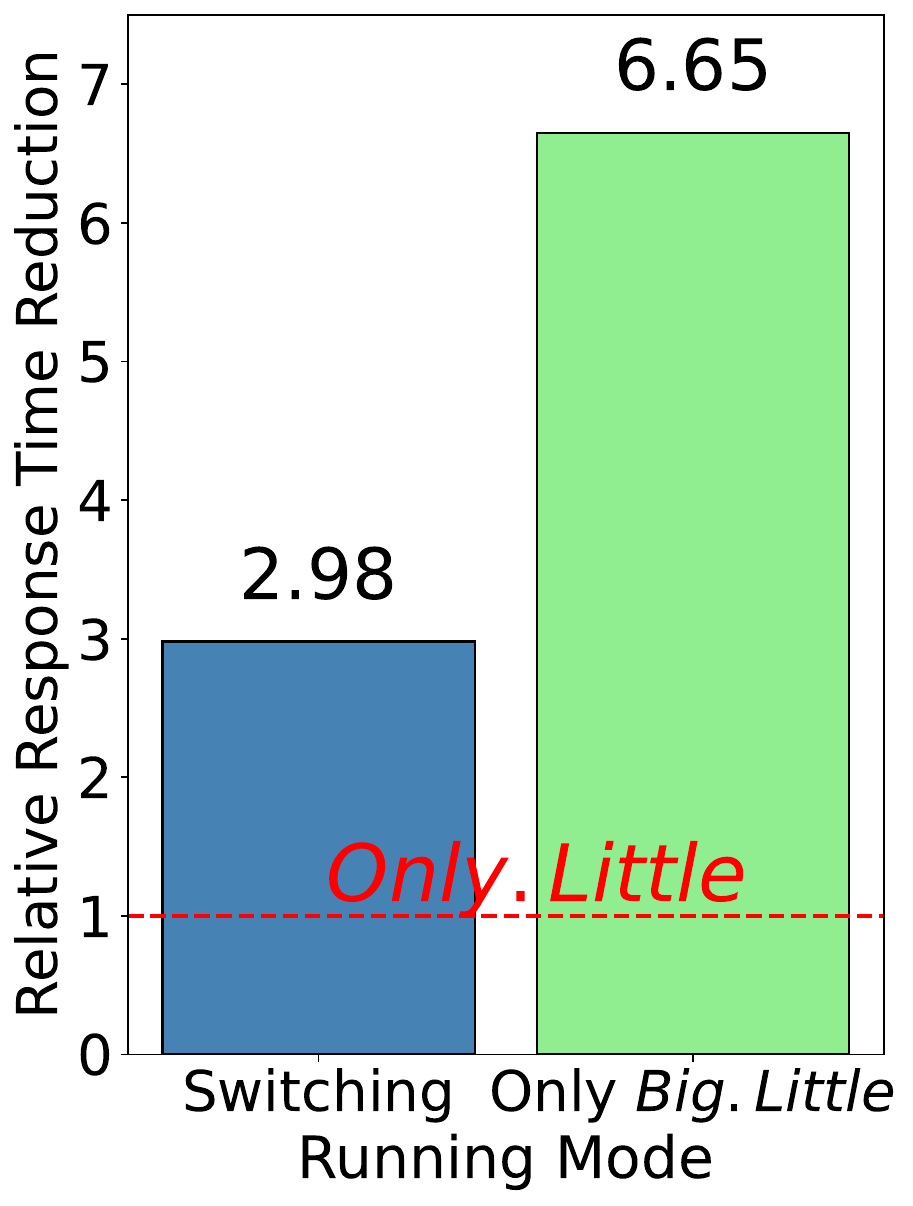}
  \end{subfigure}
  \caption{{\small$D_{\text{switch}}$} and relative response time with switching.}
  \label{fig:cross_board_switch_result}
\end{figure}
% \vspace{-3mm}
To validate the effectiveness of cross-board switching, we tested three long workloads, each with 80 applications and standard arrival intervals. Figure \ref{fig:cross_board_switch_result} (left) shows the $D_{\text{switch}}$ variation for one workload every 4 application updates, where the $D_{\text{switch}}$ value triggers a switch from \textit{Only.Little} to \textit{Big.Little} at the threshold, enabling tasks to execute on the new Big.Little FPGA. Compared to execution solely on \textit{Only.Little}, cross-board switching reduces the average response time up to nearly 3x with the average switching overhead of 1.13ms.

% \vspace{-1.5mm}
\section{Conclusion}
% \vspace{-0.5mm}
This paper presents VersaSlot, an efficient spatio-temporal FPGA sharing system with the novel \textit{Big.Little} slot architecture to address the PR contention problem. We propose a slot allocation and dual-core scheduling algorithm and a seamless cross-board switching and live migration mechanism for the architecture to maximize FPGA multiplexing in the cluster. VersaSlot achieves up to 2.19x average response time improvement over the state-of-the-art system and enhances the LUT and FF utilization by 35\% and 29\% on average.

\section*{Acknowledgment}
We extend our sincere gratitude to the Leibniz Supercomputing Centre (LRZ) and the Munich Quantum Valley (MQV) project for the support in providing FPGA resources.
\bibliographystyle{IEEEtran}
\bibliography{ref}

\end{document}